\documentclass[letterpaper, preprint, paper,11pt]{AAS}

\usepackage{bm}
\usepackage{amsmath}
\usepackage{graphicx}
\usepackage{amsfonts}
\usepackage[colorlinks=true, pdfstartview=FitV, linkcolor=black, citecolor= black, urlcolor= black]{hyperref}
\usepackage{overcite}
\usepackage{gensymb}
\usepackage{footnpag}
\usepackage[percent]{overpic}
\usepackage{caption}
\usepackage{subcaption}

\newcommand{\transp}[1]{{#1}^{\scriptscriptstyle\top}}

\PaperNumber{XX-XXX}

\begin{document}

\title{Autonomous Six-Degree-of-Freedom Spacecraft Docking Maneuvers via Reinforcement Learning}

\author{Charles E. Oestreich\thanks{Graduate Student, Department of Aeronautics and Astronautics, Massachusetts Institute of Technology, 125 Massachusetts Ave, Cambridge, MA 02139.},  
Richard Linares\thanks{Charles Stark Draper Assistant Professor, Department of Aeronautics and Astronautics, Massachusetts Institute of Technology, 125 Massachusetts Ave, Cambridge, MA 02139.},
and Ravi Gondhalekar\thanks{Senior Member of the Technical Staff, Guidance \& Control Group, The Charles Stark Draper Laboratory, Inc., 555 Technology Square, Cambridge, MA 02139.}
}

\maketitle{}

\begin{abstract}
A policy for six-degree-of-freedom docking maneuvers is developed through reinforcement learning and implemented as a feedback control law. Reinforcement learning provides a potential framework for robust, autonomous maneuvers in uncertain environments with low on-board computational cost. Specifically, proximal policy optimization is used to produce a docking policy that is valid over a portion of the six-degree-of-freedom state-space while striving to minimize performance and control costs. Experiments using the simulated Apollo transposition and docking maneuver exhibit the policy's capabilities and provide a comparison with standard optimal control techniques. Furthermore, specific challenges and work-arounds, as well as a discussion on the benefits and disadvantages of reinforcement learning for docking policies, are discussed to facilitate future research. As such, this work will serve as a foundation for further investigation of learning-based control laws for spacecraft proximity operations in uncertain environments.
\end{abstract}

\section{Introduction}
Emerging technologies such as satellite servicing and active orbital debris removal rely on autonomous docking capabilities. Current missions, including NASA's Restore-L \cite{vavrina2019restoreL} and DARPA's Robotic Servicing of Geosynchronous Satellites (RSGS) \cite{sullivan2015darparsgs}, are highly constrained and planned far in advance; it is expected that the number of such missions will increase and their scope expanded to a wide range of orbits, targets, and objectives. As such, guidance and control algorithms for proximity operations and docking will require higher levels of autonomy to deal with challenges such as varying constraints, target motion, fault-tolerance, and uncertain dynamic environments \cite{nasa2015taroadmap, starek2016survey}. 

There have been numerous research efforts to address these challenges. Lee and Pernicka \cite{lee2010optcontrol} presented an optimal control method for the Space Shuttle V-bar rendezvous trajectory with the International Space Station. Boyarko et al. \cite{boyarko2011tumble} generated minimum-time and minimum-fuel trajectories for rendezvous with a tumbling target, but the method is not implementable in real-time. Weiss et al. \cite{weiss2015mpc} demonstrated good performance for docking while abiding by path and control constraints using model predictive control; however the study was limited to three-degree-of-freedom (3-DOF) scenarios that excluded attitude dynamics and control. Jiang et al. \cite{jiang2016fault} provided actuator fault-tolerance and disturbance robustness using an adaptive fixed-time controller, but again, only the 3-DOF position was considered in docking trajectories. Jewison \cite{jewison2017uncertain} introduced a method of generating probabilistically optimal trajectories with uncertainty regarding the target spacecraft state and obstacles. Malyuta et al. \cite{malyuta2019apollo} recently developed a six-degree-of-freedom (6-DOF) docking trajectory optimization method using successive convexification that is able to address state-triggered constraints.

This research investigates reinforcement learning (RL) as an alternative solution to the aforementioned challenges. RL involves learning a policy that maps observations to actions in order to maximize a reward signal given by the environment. Since it is a general, model-free framework, RL is potentially advantageous over model-based methods for scenarios where model identification is infeasible or prohibitive, e.g., when environments, dynamics, or disturbances are time-varying. Moreover, once learned, implementation of the policy requires low amounts of computational effort and memory, making it practically realizable with current spacecraft computing resources. There have been numerous applications of RL within the robotics community for control tasks \cite{ng2003helicopter, bojarski2016selfdrive}. 

However, its extension to spacecraft guidance, navigation, and control is still largely unexplored. Recent efforts include the application of the REINFORCE algorithm for asteroid mapping (Chan and Agha-mohammadi \cite{chan2019imaging}) and the use of an RL actor-critic framework to widen the capabilities of the zero-effort-miss/zero-effort-velocity guidance algorithm, generalizing it for path constraints in near-rectilinear orbits (Scorsoglio et al. \cite{scorsoglio2019near}). Broida and Linares \cite{broida2019rendezvous} used proximal policy optimization (PPO) to accomplish 3-DOF rendezvous trajectories that exploit relative orbital dynamics. Likewise, Gaudet et al. \cite{gaudet2019mars, gaudet2019asteroid} used PPO for producing 6-DOF planetary landings and asteroid hovering maneuvers. The latter work utilized meta-learning, where the RL agent can adapt to a novel environment from learning on a wide range of possible environments. Implementing meta-learning via recurrent neural networks, the trained agent was able to adapt to unique asteroid dynamic environments and actuator faults. Finally, Hovell and Ulrich \cite{hovell2020rlguidance} recently presented a guidance policy for 3-DOF proximity operations using the ``distributed distributional deep deterministic policy gradient" (D4PG) algorithm, testing it successfully in granite surface hardware experiments. The hardware implementation distinguishes this work from most research efforts that are limited to simulation.

The research presented in this paper extends the use of RL for close-proximity approach and docking maneuvers that require 6-DOF dynamic modeling. To the best of the authors' knowledge, RL has not been applied in a truly 6-DOF scenario representative of the final docking approach between two spacecraft. In this work, PPO is used to develop a precise docking policy that is valid for initial conditions within a subset of the state-space, while also preventing collisions and minimizing control and error costs. The policy is implemented as a feedback control law. Experimental results on the simulated Apollo transposition and docking maneuver \cite{apollo1970constraints} demonstrate the policy's capabilities in a realistic 6-DOF docking scenario. This work also includes a comparison with standard optimal control methodology using the GPOPS-II software suite \cite{patterson2014gpops}. The contributions of this work are two-fold: first, to present a novel RL-based framework for 6-DOF docking policies, and, second, to provide methods, results, and insight that will benefit future research in learning-based methods for spacecraft proximity operations.

\section{Methods}
\subsection{Overview of Reinforcement Learning}
\begin{figure}[htb!]
\centering
\begin{overpic}[width=4.5in]{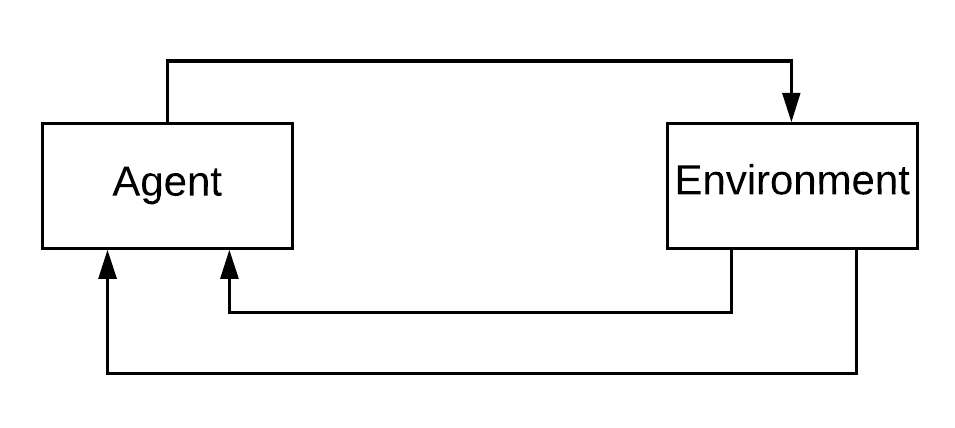}
	\put (43,39) { Action, $\mathbf{u}_t$}
	\put (43,13) { Reward, $r_t$}
	\put (34,7) {\ Next State, $\mathbf{x}_{t+1}|\mathbf{x}_t, \mathbf{u}_t$}
\end{overpic}
\caption{\textbf{The general schematic of reinforcement learning.}}
	\label{fig:rl_mdp}
\end{figure}

RL is a subdivision of machine learning where an agent learns a policy that maps observations to actions in order to maximize a numerical reward across experienced trajectories \cite{sutton2018rl}. The agent learns a policy by repeatedly interacting with the environment over numerous trajectories (termed ``episodes"), either real or simulated, and receiving rewards based on the action taken at each time step (Figure \ref{fig:rl_mdp}). RL is modeled as a Markov decision process that includes a state space $\mathcal{S}$, action space $\mathcal{A}$, state transition distribution $\mathcal{P}(\mathbf{x}_{t+1}|\mathbf{x}_t, \mathbf{u}_t)$, and reward function $r(\mathbf{x}_t, \mathbf{u}_t)$, where state $\mathbf{x} \in \mathcal{S}$, action (control input) $\mathbf{u} \in \mathcal{A}$, and $t$ is the discrete time-step index. During the learning process, the policy $\pi_{\boldsymbol{\theta}} = (\mathbf{u}_t|\mathbf{x}_t)$ is formalized as a conditional probability distribution, dependent upon the parameter vector $\boldsymbol{\theta}$, mapping states to actions.

One episode results in a trajectory of state-action pairs, denoted as $\boldsymbol{\tau}=[\mathbf{x}_0,\mathbf{u}_0,...,~\mathbf{x}_T,\mathbf{u}_T] \in \mathbb{T}$ with $T$ being the number of time steps in the trajectory and $\mathbb{T}$ being the set of all possible state-action pair trajectories. Rewards received at successive time-steps are discounted to accommodate infinite-horizon problems. The sum of discounted rewards over the trajectory is
\begin{equation}
    \label{eq:sumrewards}
    r(\boldsymbol{\tau}) = \sum_{t=0}^T\gamma^tr(\mathbf{x}_t, \mathbf{u}_t),
\end{equation}
where $\gamma \in (0, 1)$ is the discount factor. The goal of RL is to maximize the expectation of discounted rewards across all trajectories experienced by the agent:
\begin{equation}
    \label{eq:objintrewards}
    \mathbb{E}_{p_{\boldsymbol{\theta}}(\boldsymbol{\tau})}\left[r(\boldsymbol{\tau})\right] = \int_{\mathbb{T}}r(\boldsymbol{\tau})p_{\boldsymbol{\theta}}(\boldsymbol{\tau})d\boldsymbol{\tau}~~.
\end{equation}
The probability of experiencing a particular trajectory based upon the policy's parameter vector $\boldsymbol{\theta}$ is
\begin{equation}
    \label{eq:probtraj}
    p_{\boldsymbol{\theta}}(\boldsymbol{\tau})=\left[\prod_{t=0}^{T-1}p(\mathbf{x}_{t+1}|\mathbf{x}_t,\mathbf{u}_t)\right]p(\mathbf{x}_0),
\end{equation}
where $\mathbf{u}_t$ is sampled from $\pi_{\boldsymbol{\theta}}(\mathbf{u}_t|\mathbf{x}_t)$. Note that the state transition is stochastic in the general case: this can be replaced by a deterministic state transition in certain applications. For example, the Apollo docking scenario presented in this paper assumes nominal dynamics and thus a deterministic state transition is used. The variance of the policy's conditional distribution results in a stochastic action choice, enabling further exploration of the action space. As learning progresses, the variance is reduced to instead encourage more exploitation of the current policy. After the learning process is completed, the variance of the policy is set to zero, resulting in a deterministic action choice (the mean value of $\pi_{\boldsymbol{\theta}}(\mathbf{u}_t|\mathbf{x}_t)$), i.e., a deterministic feedback control law during implementation.

\subsection{Proximal Policy Optimization}
The specific RL algorithm used in this research is PPO \cite{schulman2017ppo}. PPO is a state-of-the-art policy learning algorithm with successful results in many control tasks with continuous or discrete state/action spaces. PPO is a model-free, actor-critic algorithm where the policy that selects actions (the actor) and an advantage function that evaluates the selected actions (the critic) are learned concurrently. The state-value function $V_{\mathbf{w}}^{\pi}(\mathbf{x}_t)$ (with the parameter vector $\mathbf{w}$) is used in PPO and estimates the sum of future discounted rewards over the trajectory starting at the current state $\mathbf{x}_t$ and following the current policy. However, this function is initially unknown and the parameter vector $\mathbf{w}$ must be learned concurrently with the policy parameter vector $\boldsymbol{\theta}$.  The resulting advantage function $A_{\mathbf{w}}^{\pi}(\mathbf{x}_t, \mathbf{u}_t)$ is the difference between the empirical rewards received during the learning process and the state-value function's estimate.
\begin{equation}
    \label{eq:state_value}
    V_{\mathbf{w}}^{\pi}(\mathbf{x}_t)=\mathbb{E}_{\pi}\left[\sum_{k=t}^T\gamma^{k-t}r_k(\mathbf{x}_k,\mathbf{u}_k)\middle|\mathbf{x}_t\right]
\end{equation}
\begin{equation}
    \label{eq:advantage}
    A^{\pi}_{\mathbf{w}}(\mathbf{x}_t,\mathbf{u}_t) =  \left[\sum_{k=t}^T\gamma^{k-t}r_k(\mathbf{x}_k, \mathbf{u}_k)\right] - V_{\mathbf{w}}^{\pi}(\mathbf{x}_t).
\end{equation}

PPO is a descendant of the trust region policy optimization algorithm \cite{schulman2015TRPO}, retaining the ability to mitigate large policy updates (thus reducing the risk of learning divergence) while being simpler and more widely implementable. Central to PPO is the policy probability ratio \begin{equation}
    \label{eq:probratio}
    p_t(\boldsymbol{\theta})=\frac{\pi_{\boldsymbol{\theta}}(\mathbf{u}_t|\mathbf{x}_t)}{\hat{\pi}_{\boldsymbol{\theta}}(\mathbf{u}_t|\mathbf{x}_t)},
\end{equation}
which compares the probability $\pi_{\boldsymbol{\theta}}(\mathbf{u}_t|\mathbf{x}_t)$ of selecting a particular action \textit{after} a learning update to the probability $\hat{\pi}_{\boldsymbol{\theta}}(\mathbf{u}_t|\mathbf{x}_t)$ of selecting the same action \textit{prior} to the update. The probability ratio is then directly used in the PPO objective function we wish to maximize:
\begin{equation}
\label{eq:ppoobjective}
    J(\boldsymbol{\theta})=\mathbb{E}_{p(\boldsymbol{\tau})}\bigg[~\text{min}~\Big(p_t(\boldsymbol{\theta})A^{\pi}_{\mathbf{w}}(\mathbf{x}_t,\mathbf{u}_t),~ \text{clip}\big[p_t(\boldsymbol{\theta}),\epsilon\big]A^{\pi}_{\mathbf{w}}(\mathbf{x}_t,\mathbf{u}_t)\Big)\bigg]
\end{equation}
where the clip function, defined as
\begin{equation}
\mbox{clip}\big[p_t(\boldsymbol{\theta}), \epsilon\big] = 
\begin{cases}
    1 - \epsilon & \mbox{if}~ p_t(\boldsymbol{\theta}) < 1 - \epsilon
    \\ 
    1 + \epsilon & \mbox{if}~p_t(\boldsymbol{\theta}) > 1 + \epsilon \\
    p_t(\boldsymbol{\theta}) & \mbox{otherwise}
\end{cases}~~~,
\end{equation}
imposes bounds on the policy probability ratio using the clipping parameter $\epsilon \in (0,1)$. The clipping parameter controls how close the updated policy is to the old policy, effectively implementing a trust region and eliminating large, unwanted policy updates. Note that this objective function is measured relative to the policy prior to the update. Thus, the numerical value of the objective function over the course of many updates is uninformative. Instead, its immediate gradient is more critical in guiding the policy to maximize rewards over all trajectories.

To learn the state-value function, we minimize the commonly used mean squared error cost function:
\begin{equation}
    \label{eq:valueloss}
    L(\mathbf{w}) = \frac{1}{2}~\mathbb{E}_{p(\boldsymbol{\tau})} \left[\left(V_{\mathbf{w}}^{\pi}\left(\mathbf{x}_t\right) - \sum_{k=t}^{T}\gamma^{k-t}r\left(\mathbf{x}_k, \mathbf{u}_k\right)\right)^2\right].
\end{equation}
 Essentially, the mean squared difference between the state-value function estimate and the actual sum of resulting rewards is minimized. The multiplication by $\frac{1}{2}$ simplifies the loss gradient calculation. We now have an objective function to improve the policy in Equation \eqref{eq:ppoobjective} and a cost function to correct errors in the state-value function in Equation \eqref{eq:valueloss}. Thus, we can use the gradients of these functions to perform gradient ascent on $\boldsymbol{\theta}$ and gradient descent on $\mathbf{w}$:
\begin{subequations}
\label{eq:graddescent}
\begin{align}
    \boldsymbol{\theta}^+&=\boldsymbol{\theta}^-+ \beta_{\boldsymbol{\theta}}\nabla_{\boldsymbol{\theta}}J(\boldsymbol{\theta})|_{\boldsymbol{\theta}=\boldsymbol{\theta}^-} \\
    \mathbf{w}^+&=\mathbf{w}^-- \beta_{\mathbf{w}}\nabla_{\mathbf{w}}L(\mathbf{w})|_{\mathbf{w}=\mathbf{w^-}}
\end{align}
\end{subequations}
where the scalars $\beta_{\boldsymbol{\theta}}$ and $\beta_{\mathbf{w}}$ are the policy learning rate and state-value function learning rate, respectively, which must be chosen by the designer.

\subsection{Implementation for 6-DOF Docking}
For developing policies to perform 6-DOF docking maneuvers, we use the target-centered inertial frame $\mathcal{I}$ (assuming the target is stationary) and the chaser-centered body frame $\mathcal{B}$. The state $\mathbf{x}=\transp{[\transp{\mathbf{r}},~\transp{\mathbf{v}},~\transp{\mathbf{q}},~\transp{\boldsymbol{\omega}}]}$ is the state of the chaser's center of mass within the $\mathcal{I}$ frame, with $\mathbf{r} \in \mathbb{R}^3$ as the position, $\mathbf{v} \in \mathbb{R}^3$ as the velocity, $\mathbf{q} \in \mathbb{R}^4$ as the attitude quaternion, and $\boldsymbol{\omega} \in \mathbb{R}^3$ as the angular velocity. The control action $\mathbf{u}=\transp{[\transp{\mathbf{F}},~ \transp{\mathbf{L}}]}$ consists of a thrust command $\mathbf{F} \in \mathbb{R}^3$ and a torque command $\mathbf{L} \in \mathbb{R}^3$, both in the $\mathcal{I}$ frame. The thrust and torque commands are both bounded by minimum/maximum actuator constraints. Control actions are commanded by the policy at discrete time intervals. The dynamics are derived below in continuous-time, and are subsequently discretized using a sample-period of 1 second.

The translational dynamics are modeled using the double-integrator equations: 
\begin{subequations}
\label{eq:doubleintegrate}
\begin{align}
    \dot{\mathbf{r}} &= \mathbf{v} \\
    \dot{\mathbf{v}} &= \frac{\mathbf{F}}{m} 
\end{align}
\end{subequations}
where $m$ refers to the chaser mass. In practice, to compute individual thruster commands, the net force command must be determined in the chaser body frame. This is calculated using the chaser's current attitude, parameterized as a rotation matrix $\mathbf{R}(\mathbf{q}) \in \mathbb{R}^{3\times3}$, that maps from the $\mathcal{B}$ frame to the $\mathcal{I}$ frame:
\begin{equation}
    \label{eq:f2body}
    \mathbf{F}_{\mathcal{B}} = \transp{\mathbf{R}(\mathbf{q})}\mathbf{F}.
\end{equation}

The attitude dynamics are modeled using quaternion kinematics and Euler's equations for rigid bodies:
\begin{subequations}
\label{eq:eulers}
\begin{align}
    \dot{\mathbf{q}} &= \frac{1}{2}\boldsymbol{\Omega}\boldsymbol{\omega} \\
    \dot{\boldsymbol{\omega}} &= \mathbf{J}^{-1}\left(\mathbf{L} - \boldsymbol{\omega} \times \mathbf{J}\boldsymbol{\omega}\right) 
\end{align}
\end{subequations}
where $\mathbf{J}$ is the chaser inertia tensor and $\boldsymbol{\Omega}$ is defined as 
\begin{equation}
    \boldsymbol{\Omega} = \begin{bmatrix} -q_x & -q_y & -q_z \\ q_w & -q_z & q_y \\ q_z & q_w & -q_x \\ -q_y & q_x & q_w \end{bmatrix}~~.
\end{equation}

Thus, Equations \eqref{eq:doubleintegrate} and \eqref{eq:eulers} govern the dynamics of the nonlinear, 6-DOF system. As docking requirements are often formulated on the relative state of the chaser and target docking ports, it is useful to define the relative position $\mathbf{r}_p$ and velocity $\mathbf{v}_p$ of the chaser docking port with respect to the target docking port as
\begin{subequations}
\label{eq:docking_ports}
\begin{align}
    \mathbf{r}_p &= \mathbf{r} + \mathbf{R}(\mathbf{q})\mathbf{r}_c - \mathbf{r}_t \\
    \mathbf{v}_p &= \mathbf{v} + \left(\boldsymbol{\omega} \times \mathbf{R}(\mathbf{q})\mathbf{r}_c\right)
\end{align}
\end{subequations}
where $\mathbf{r}_c$ and $\mathbf{r}_t$ refer to the chaser and target docking port positions in the $\mathcal{B}$ and $\mathcal{I}$ frames, respectively. Note that, if the target is stationary in the $\mathcal{I}$ frame, the relative attitude and angular velocity of the chaser docking port is equivalent to the chaser's attitude and angular velocity.

The policy and state-value function are modeled through standard, feedforward neural networks. Thus, the parameters $\boldsymbol{\theta}$ and $\mathbf{w}$ represent the weights and biases of each network's respective layers. The neural network backpropagation algorithm and Adam optimizer\cite{adam2014} are used to perform gradient ascent/descent on the parameter vectors $\boldsymbol{\theta}$ and $\mathbf{w}$ according to Equation \eqref{eq:graddescent}. The policy is specifically a multivariate, Gaussian distribution with a diagonal covariance matrix. The neural network output consists of the resulting mean action based on the given state. The variance for each action is also learned, but is independent of the state. This essentially controls the degree of action space exploration throughout the learning process. In general, the agent learns to set large variance values during the learning process to encourage more exploration, and then diminish them in the later stages of learning to induce more exploitation of the current policy.

The inputs for both the policy and state-value function neural networks are scaled using a running mean and standard deviation of experienced state data while learning. This helps prevent the saturation of activation functions within each network layer. Similarly, neural network outputs are best defined when close to unity. As such, the policy network outputs are scaled accordingly so that an output of $\pm 1$ corresponds to the maximum/minimum thrust or torque command. Table \ref{nnparams} shares the structure of network layers and their corresponding activation functions.
\begin{table}[htb!]
	\fontsize{10}{10}\selectfont
    \caption{Neural network structural parameters.}
   \label{nnparams}
        \centering 
   \begin{tabular}{c | c | c | c | c } 
      \hline 
          & \multicolumn{2}{c|}{Policy Network} & \multicolumn{2}{c}{State-Value Function Network} \\
      \hline 
      Layer & Neurons & Activation & Neurons & Activation \\
      \hline
      1st hidden & 130 & tanh & 130 & tanh \\
      2nd hidden & 88  & tanh & 25  & tanh \\
      3rd hidden & 60  & tanh & 5   & tanh \\
      Output     & 6   & linear & 1   & linear \\
      \hline
   \end{tabular}
\end{table}

In our implementation of PPO, we follow the example of Gaudet et al. \cite{gaudet2019mars} in that we dynamically adjust learning parameters to target a desired Kullback-Leibler (KL) divergence value between successive policy updates \cite{kullback1951KLDiv}. This helps to prevent large policy updates that could derail the learning process, while yielding smoother policy updates. Over the course of learning, both the PPO clipping parameter $\epsilon$ and the policy learning rate $\beta_{\boldsymbol{\theta}}$ are adjusted to keep the KL-divergence between updates as close as possible to the desired target value ($KL_{\mathrm{des}}$).

Employing a valid reward function is critical to the success of PPO as the policy will learn to explicitly maximize this function. For 6-DOF docking maneuvers, the reward function consists of several terms that together account for minimizing state tracking errors and control effort, preventing collisions, and reinforcing successful docks. All terms are weighted relative to one another through design coefficients.

To account for translational state error, the term $\transp{(\dot{\mathbf{v}}_t - \dot{\bar{\mathbf{v}}}_t)}\mathbf{M}(\dot{\mathbf{v}}_t - \dot{\bar{\mathbf{v}}}_t)$  defines the quadratic weighted error between the acceleration produced by the RL agent via Equation (\ref{eq:doubleintegrate}b) and a reference acceleration ($\dot{\bar{\mathbf{v}}}_t$) provided by a linear quadratic regulator (LQR) feedback law:
\begin{equation}
\label{eq:lqr}
    \dot{\bar{\mathbf{v}}}_t = -\mathbf{K}\mathbf{x}'_t
\end{equation}
where $\mathbf{K}$ is the LQR gain matrix and $\mathbf{x}'_t$ is the translational part of the state vector (chaser position and velocity). The LQR design process is performed offline before the implementation of PPO. By adjusting the standard LQR performance and control cost matrices, the resulting gain matrix can be tuned to target a desired trajectory time length for nominal docking initial conditions. Additionally, the tuning process can account for a desired, non-zero final docking velocity by setting the LQR origin at an offset from the actual docking port location. The benefits of this reward term are two-fold: first, it provides a clear reward signal at all points in the translational state-space that guides the RL agent to achieve a successful docking trajectory, and, second, it encourages the RL agent to produce docking trajectories with a specific time length (an important design consideration for many docking maneuvers). Finally, $\mathbf{M} \in \mathbb{R}^{3\times3}, \mathbf{M} \succ \mathbf{0}$ where the weights in $\mathbf{M}$ are design parameters.

To account for errors between the actual and desired attitude and angular velocity, we define the reward function term $\transp{\widetilde{{\boldsymbol{\alpha}}}}_t\mathbf{Q}\widetilde{{\boldsymbol{\alpha}}}_t$. This term is based on the error quaternion $\mathbf{\widetilde{q}}$ between the current and desired final docking attitude ($\mathbf{q}_{\mathrm{des}}$):
\begin{equation}
\label{eq:quaterror}
    \mathbf{\widetilde{q}}=\mathbf{q}_{\mathrm{des}}\otimes\mathbf{q}^{-1}~~.
\end{equation}
From Markley \cite{markley2003error}, we take twice the vector component of the error quaternion ($\widetilde{\mathbf{q}}_v$) as our measure of attitude error. We also penalize the angular velocity error $\widetilde{\boldsymbol{\omega}}$ between the current angular velocity and the desired angular velocity $\boldsymbol{\omega}_{\mathrm{des}}$. This results in a vector $\widetilde{\boldsymbol{\alpha}}_t = \transp{[2\transp{\widetilde{\mathbf{q}}}_v,~\transp{\widetilde{\boldsymbol{\omega}}}]} \in \mathbb{R}^6$ that penalizes both attitude and angular velocity errors. This term also has a weighting design matrix $\mathbf{Q} \in \mathbb{R}^{6\times6}, \mathbf{Q} \succ \mathbf{0}$.

Finally, we include a quadratic control cost, collision penalty, and docking bonus. The quadratic control cost $\transp{\mathbf{u}}_t\mathbf{P}\mathbf{u}_t$ is applied to the force/torque command with a weighting design matrix $\mathbf{P} \in \mathbb{R}^{6\times6}, \mathbf{P} \succ \mathbf{0}$. The collision penalty $c\sin\left({\frac{\pi}{2}\frac{||\mathbf{r}_p||}{r_{\mathrm{col}}}}\right)$ is applied at each time step the chaser docking port is within the boundary of the target spacecraft (modeled as a rectangular region). The penalty is scaled based upon the relative docking distance at the time of collision and the maximum possible distance for a collision ($r_{\mathrm{col}}$). Together with the sine function and a weighting coefficient $c > 0$, this permits a smooth function that penalizes collisions with significant state error more heavily. The docking bonus $g(\mathbf{x}_t)$ is a discrete term applied if the agent achieves the docking requirements:
\begin{equation}
g(\mathbf{x}_t) = 
\begin{cases}
    d & \mbox{if $\mathbf{x}_t$ satisfies docking conditions} \\ 0 & \mbox{otherwise}
\end{cases}
\end{equation}
where $d > 0$ is a weighting coefficient.

We define two distinct reward functions based on the aforementioned reward contributions:
\begin{subequations}
\label{eq:tworewards}
\begin{align}
    r_1(\mathbf{x}_t,\mathbf{u}_t)&=-\transp{(\dot{\mathbf{v}}_t - \dot{\bar{\mathbf{v}}}_t)}\mathbf{M}(\dot{\mathbf{v}}_t - \dot{\bar{\mathbf{v}}}_t)-\transp{\widetilde{{\boldsymbol{\alpha}}}}_t\mathbf{Q}\widetilde{{\boldsymbol{\alpha}}}_t-\transp{\mathbf{u}}_t\mathbf{P}\mathbf{u}_t-c\sin\left({\frac{\pi}{2}\frac{||\mathbf{r}_p||}{r_{\mathrm{col}}}}\right) \\
    r_2(\mathbf{x}_t) &= g(\mathbf{x}_t)
\end{align}
\end{subequations}
where $r_1$ represents the ``shaping" penalties (LQR error, attitude/angular velocity error, control cost, and collision penalty) and $r_2$ represents the terminal docking bonus. Following the example of Gaudet et al. \cite{gaudet2019mars}, a slightly smaller discount factor is used for the shaping penalties ($\gamma_1$) while a larger discount factor is used for the terminal docking bonus ($\gamma_2$). This results in the docking bonus being weighted more heavily in the long-term than the shaping penalties. Thus, the advantage function (Equation \eqref{eq:advantage}) and the state-value loss function (Equation \eqref{eq:valueloss}) can be re-written as:
\begin{equation}
\label{eq:doublediscountadv}
    A^{\pi}_{\mathbf{w}}(\mathbf{x}_t,\mathbf{u}_t)= \sum_{k=t}^T\left[\gamma_1^{k-t}r_1(\mathbf{x}_k, \mathbf{u}_k)+\gamma_2^{k-t}r_2(\mathbf{x}_k)\right] - V_{\mathbf{w}}^{\pi}(\mathbf{x}_t)
\end{equation}
\begin{equation}
\label{eq:doublediscountloss}
    L(\mathbf{w}) = \frac{1}{2}~\mathbb{E}_{p(\boldsymbol{\tau})} \left[\left(V_{\mathbf{w}}^{\pi}\left(\mathbf{x}_t\right) - \sum_{k=t}^{T}\left[\gamma_1^{k-t}r_1\left(\mathbf{x}_k, \mathbf{u}_k\right)+\gamma_2^{k-t}r_2\left(\mathbf{x}_k,\mathbf{u}_k\right)\right]\right)^2\right].
\end{equation}
The training episode either ends if the docking requirements are met or a trajectory time limit has been reached.

\section{Experimental Setup}
\begin{figure}[htb!]
	\centering\includegraphics[width=4.5in]{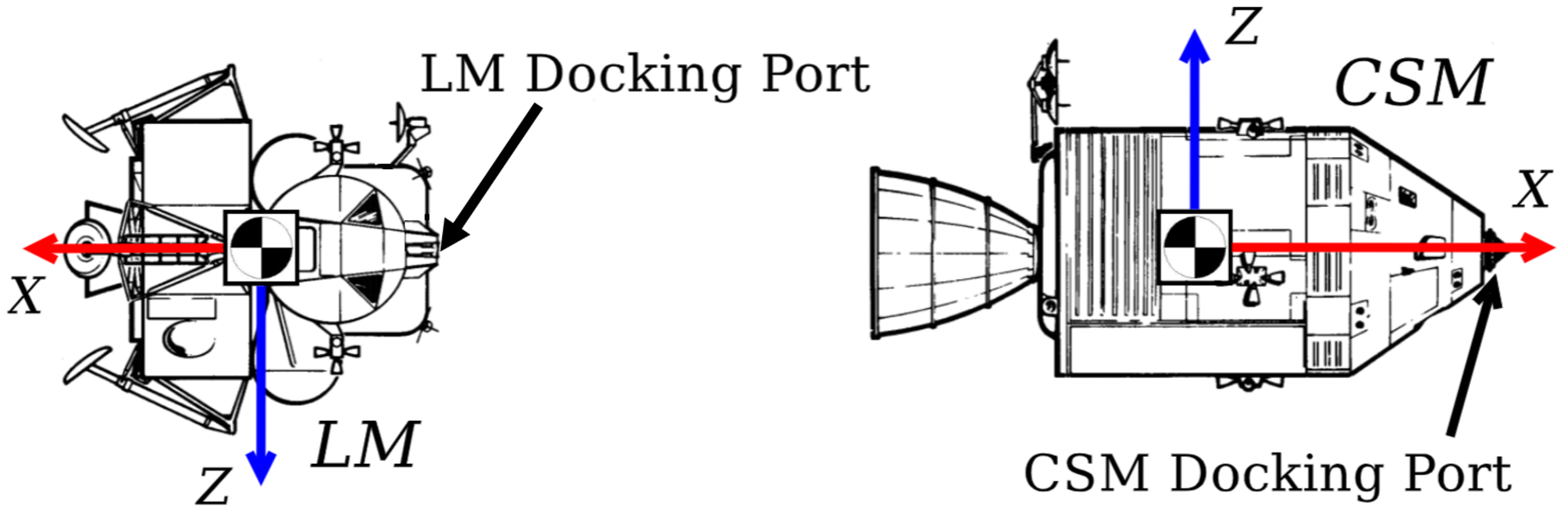}
	\caption{\textbf{Overview of coordinate frames and initial configuration for the Apollo transposition and docking maneuver. The $Y$-axis is defined according to the right-hand rule. LM/CSM diagram credit: NASA.}}
	\label{figframes}
\end{figure}

The above methodology was applied to the simulated Apollo transposition and docking maneuver. This maneuver involves the command service module (CSM) re-orienting itself and docking with the lunar module (LM) in order to extract it from the third stage of the Saturn V rocket \cite{apollo1969massprop}. Figure \ref{figframes} depicts the relevant coordinate frames: the $LM$ frame represents the target-centered inertial frame (situated at the LM's center of mass and coinciding with the LM's body frame), while the $CSM$ frame represents the chaser's body frame (situated at the CSM's center of mass). Note that Figure \ref{figframes} shows the LM and CSM in their initial configuration for the maneuver: the $CSM$ frame has a 180\degree~pitch rotation relative to the $LM$ frame.  

\begin{figure}[htb!]
	\centering\includegraphics[width=5.0in]{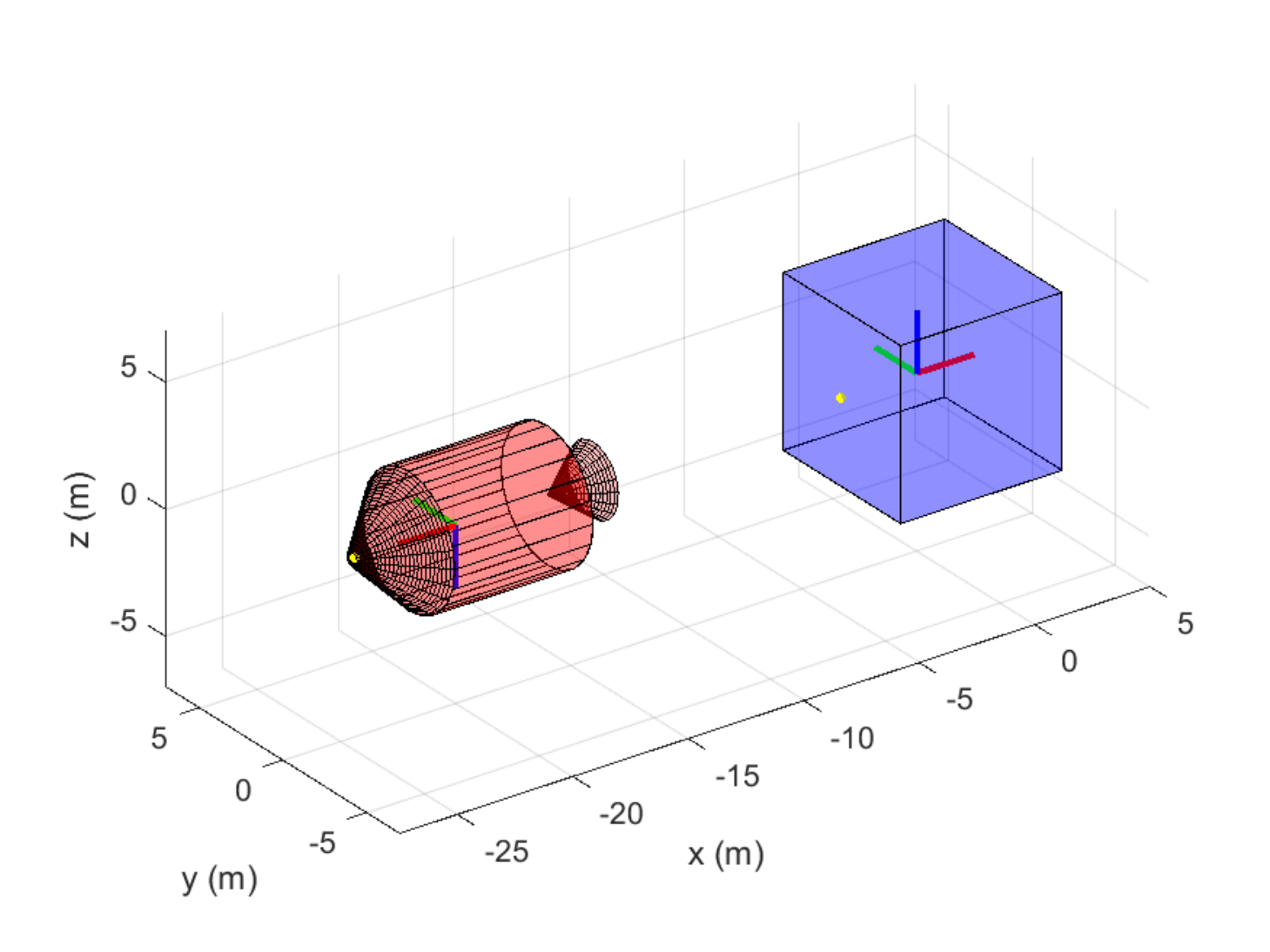}
	\caption{\textbf{Modeling the Apollo transposition and docking maneuver. The CSM is shown in red while the LM rectangular collision area is shown in blue. The LM docking port is the yellow point on the collision area surface.}}
	\label{figcollision}
\end{figure}

We impose actuator constraints on the minimum/maximum force and torque commands (Table \ref{thrusttable}). These are derived from approximating the maximum thrust and torque outputs achievable through the CSM's sixteen reaction control system thrusters \cite{apollo1969rcs}. Note that the policy produces thrust and torque commands within a continuous range of values; in reality, the CSM thrusters produce discrete, on/off thrust. We also define a rectangular region around the LM's center of mass to model collisions (Figure \ref{figcollision}). The region's $y$ and $z$ dimensions approximate the largest diameter of the LM and its surface intersects with the LM docking port. The collision geometry parameters, as well as the docking port positions of the CSM/LM in their respective spacecraft frames, are shown in Table~\ref{tablegeometry}. To perform a successful dock, the CSM docking port must meet the conditions outlined in Table \ref{tabledockconditions} ($\phi, \theta, \psi$ represent the Euler attitude angles in the $x$, $y$, and $z$ axes, respectively) \cite{apollo1970constraints}. Notice that the maneuver requires a strictly positive velocity and a $-60$\degree~roll in the $x$-axis to activate the docking mechanism.
\begin{table}[htb!]
	\fontsize{10}{10}\selectfont
    \caption{Actuator minimum/maximum constraints.}
   \label{thrusttable}
        \centering 
   \begin{tabular}{c | c | c  } 
      \hline 
      Action Command  & Value & Units \\
      \hline 
      $\mathbf{F}$ & $\pm~[790.80,~790.80,~790.80]$ & N  \\
      $\mathbf{L}$ & $\pm~[2534.91,~2534.91,~2534.91]$ & N-m  \\
      \hline
   \end{tabular}
\end{table}
\begin{table}[htb!]
	\fontsize{10}{10}\selectfont
    \caption{Collision geometry and docking port parameters.}
   \label{tablegeometry}
        \centering 
   \begin{tabular}{c | c | c  } 
      \hline 
      Parameter  & Value & Units \\
      \hline 
      Collision box $y$-dim. & $7$ & m  \\
      Collision box $z$-dim. & $7$ & m \\
      $\mathbf{r}_c$ & $[4.479, 0, 0]$ & m \\
      $\mathbf{r}_t$ & $[-3.250, 0, 0]$ & m \\
      \hline
   \end{tabular}
\end{table}
\begin{table}[htb!]
	\fontsize{10}{10}\selectfont
    \caption{Conditions for successful docking.}
   \label{tabledockconditions}
        \centering 
   \begin{tabular}{c | c | c | c  } 
      \hline 
         Docking State Term & Goal & Acceptable Deviation  & Units \\
      \hline 
      $\mathbf{r}_p$ & $\mathbf{0}$ & ~$\pm~0.15$ & m \\
      $v_{px}$ & $0.1$ & $[0.05,~0.15]$ & m/s \\
      $v_{py},~v_{pz}$ & $0,~0$ & ~$\pm~0.1$ & m/s \\
      $\phi,~\theta,~\psi$ & $-60,~0,~0$ & ~$\pm~5$ & deg \\
      $\boldsymbol{\omega}$ & $\mathbf{0}$& ~$\pm~0.75$ & deg/s \\
      \hline
   \end{tabular}
\end{table}

For both training and testing episodes, the policy accepts the given state and produces a commanded force/torque at discrete, 1-second intervals. Episodes during training are limited to 150 seconds (a rough approximation of the time needed to dock) to gather suitable data while retaining efficiency in the overall learning process. However, for testing, the time limit is extended to 250 seconds (an arbitrary time limit greater than any time length needed for successful docking) to make sure valid docking trajectories are not prematurely terminated. 

An objective of this research is to synthesize a feedback control law that is robust to significant uncertainty in the initial condition of the docking maneuver. To this end, the RL goal is to generate a docking policy that can successfully be employed within a wide range of initial conditions. This range of initial conditions should wholly contain the range of uncertainty with respect to which robustness is required. For the Apollo transposition and docking maneuver, we specifically define the initial condition range shown in Table \ref{ictable}. Monte Carlo tests of the policy randomly sample an initial condition (in each state variable) for the trajectory according the ``Testing Range". However, to ensure the policy learns across a wide region of the state space, and thus gains more robust qualities, each training episode's initial conditions are sampled from the wider ``Training Range" (also shown in Table \ref{ictable}).

\begin{table}[htb!]
	\fontsize{10}{10}\selectfont
    \caption{Initial condition range. The testing range is used for closed-loop simulation testing while the training range is employed during learning.}
   \label{ictable}
        \centering 
      \begin{tabular}{c|c|c|c}
      \hline
      & Testing Range & Training Range & Units\\
      \hline
      $\mathbf{r}$ & $[-20,~0,~0]~\pm~2$ & $[-20,~0,~0]~\pm~4$ & m\\
      $\mathbf{v}$ & $[0, 0, 0]~\pm~0.1$ & $[0, 0, 0]~\pm~0.2$ & m/s \\
      $[\phi,~\theta,~\psi]$ & $[0, 180, 0]~\pm~20$ & $[0, 180, 0]~\pm~40$ & deg\\
      $\boldsymbol{\omega}$ & $[0,~0,~0]~\pm~5$ & $[0,~0,~0]~\pm~10$ & deg/s\\
 \hline
   \end{tabular}
\end{table}

\begin{figure}[htb!]
\begin{subfigure}{.45\textwidth}
\centering
\includegraphics[width=1.0\linewidth]{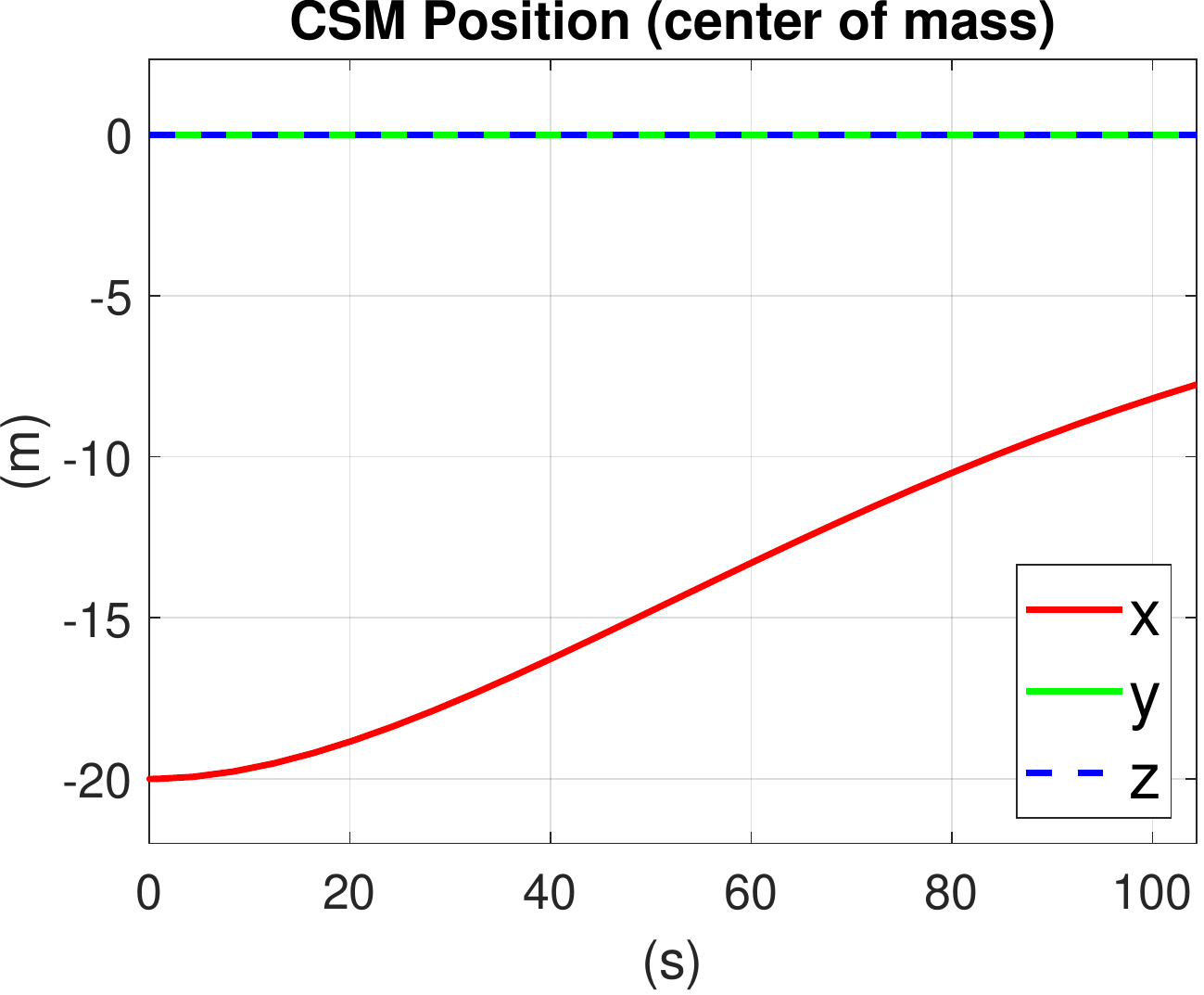}
\end{subfigure} \hfill%
\begin{subfigure}{.45\textwidth}
\centering
\includegraphics[width=1.0\linewidth]{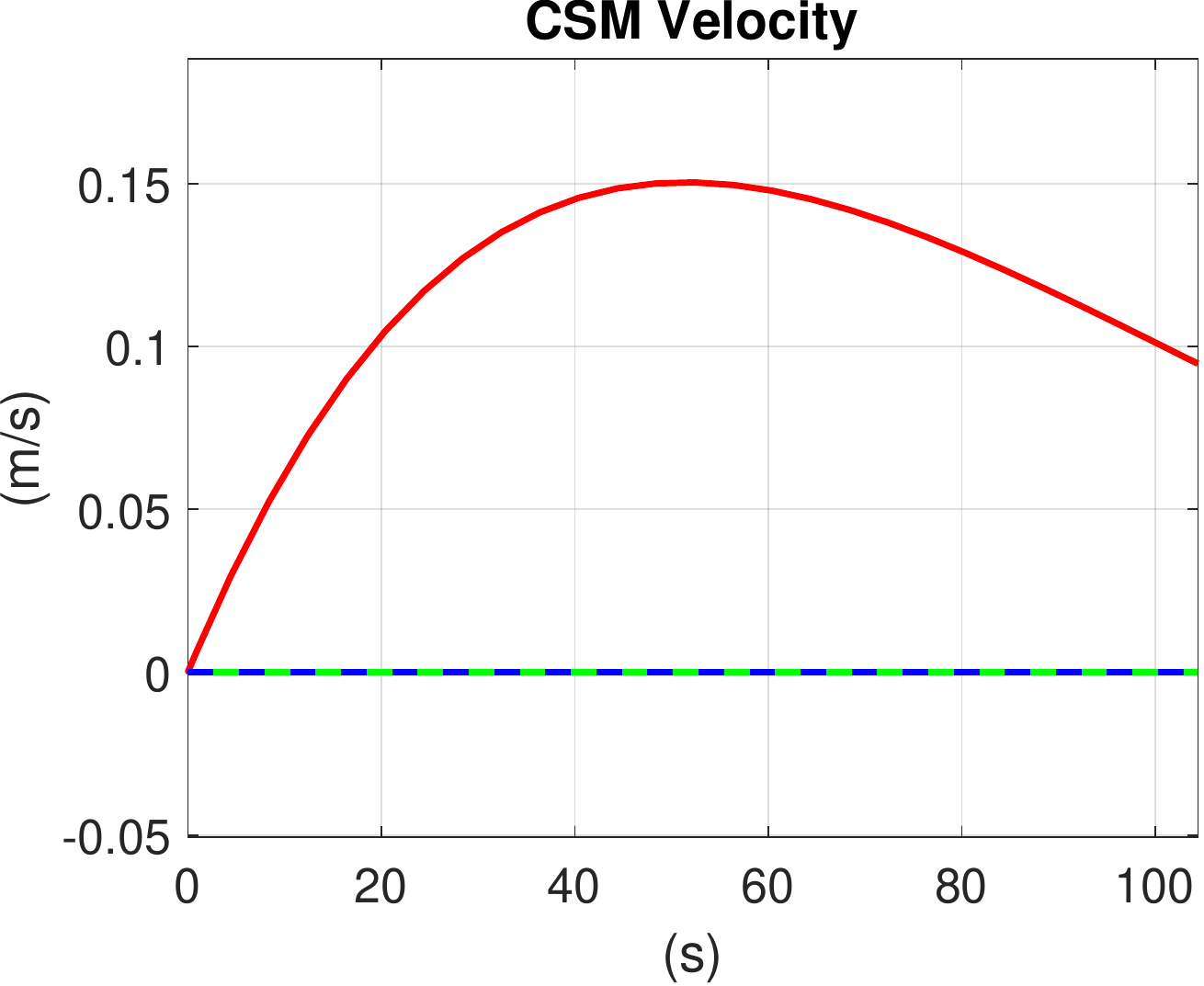}
\end{subfigure}
\vspace{5mm}
	\caption{\textbf{The LQR reference for the translational trajectory in the nominal initial condition case. Note that the final CSM center of mass position corresponds to the co-location of the CSM/LM docking ports at the correct attitude (see Table \ref{tablegeometry} for the CSM/LM docking port locations within their respective spacecraft frames).}}
	\label{fig:lqr_ref}
\end{figure}

The LQR reference gain $\mathbf{K}$ was calculated by tuning the LQR cost terms to result in a translational trajectory lasting roughly 105 seconds for the nominal initial condition case ($\mathbf{r}_0 = [-20,~0,~0]$ m, $\mathbf{v}_0 = [0,~0,~0]$ m/s). The LQR origin state was adjusted by a +3 meter offset in the $x$-axis to account for the required non-zero final velocity. The resulting position and velocity of CSM center of mass from the LQR reference accelerations is shown in Figure \ref{fig:lqr_ref}.

\section{Results}
\subsection{Learning Results}
For training the agent to perform the Apollo transposition and docking maneuver, we implemented PPO using PyTorch\footnote{https://pytorch.org/} and building on Patrick Coady's open-source work\footnote{https://github.com/pat-coady/trpo}. Learning occurred over 600,000 episodes. The agent accumulated batches of 128 episodes before performing a policy and state-value function learning update (according to Equation \eqref{eq:graddescent}) using the collected data. The initial conditions for each episode were randomly sampled from the training range shown in Table \ref{ictable}. Table \ref{trainingtable} shares key learning parameters.

\begin{table}[htb!]
	\fontsize{10}{10}\selectfont
    \caption{Training and Reward Parameters}
   \label{trainingtable}
        \centering 
      \begin{tabular}{c|c}
      \hline
      $KL_{\mathrm{des}}$ & $0.001$ \\ \hline
      $\gamma_1$ & $0.98$ \\ \hline
      $\gamma_2$ & $0.995$ \\ \hline
      $\mathbf{M}$ & $\mbox{diag}\left([2,2,2]\times10^5\right)$ \\ \hline
      $\mathbf{Q}$ & $\mbox{diag}\left([20, 20, 20, 20, 20, 20]\right)$ \\ \hline
      $\mathbf{P}$ & $\mbox{diag}\left([10,10,10,1.11,1.11,1.11]\times10^{-6}\right)$ \\ \hline
      $c$ & $10$ \\ \hline
      $d$ & $1000$ \\ \hline
   \end{tabular}
\end{table}

The algorithm kept track of the ``best'' policy over the course of learning by performing a 128-episode test using the current policy after each update.  The 128 episodes for this deterministic test (i.e., no policy variance) used initial conditions at the limits of the testing ranges shown in Table \ref{ictable}. Specifically, the combinations of the minimum/maximum initial values for the seven variables $r_x$, $v_x$, $v_y$, $v_z$, $\omega_x$, $\omega_y$, $\omega_z$ were used as edge cases to provide an accurate evaluation of the policy ($2^7 = 128$). If the current policy was able to achieve greater than or equal to the number of successful docks from the previous, ``best'' policy, it was saved as the new ``best'' policy. This process ensured that a high-performing policy was extracted over the duration of the learning process.

\begin{figure}[htb!]
	\centering
	\includegraphics[width=4.0in]{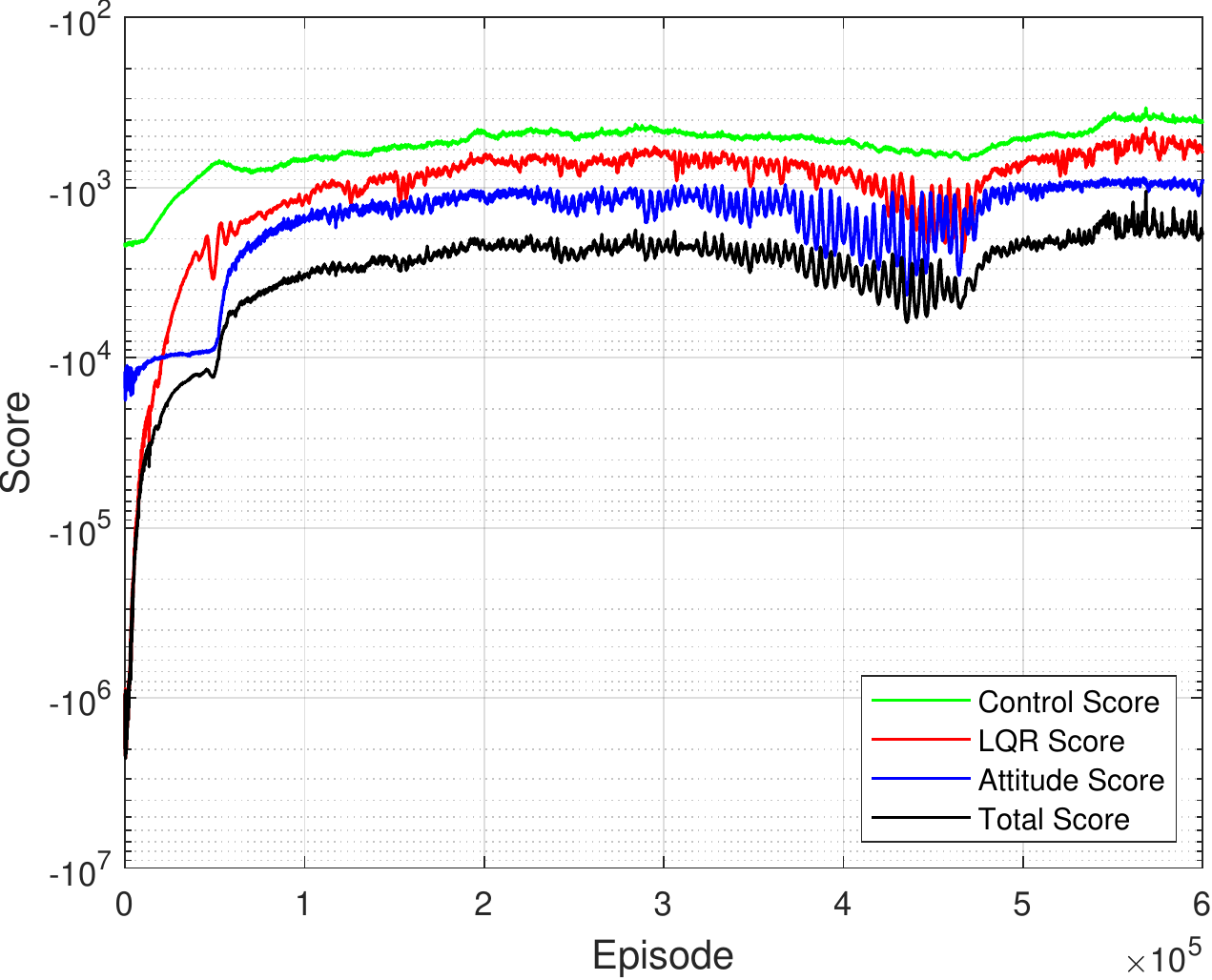}
	\caption{\textbf{Mean scores (sum of rewards) experienced per training batch over the learning process, including score portions from individual reward terms.}}
	\label{fig:rewards}
\end{figure}

Figure \ref{fig:rewards} depicts the mean score (sum of rewards) per batch over the course of the entire learning process. Also included are the score portions attributed to the main shaping reward terms in Equation \eqref{eq:tworewards} (LQR reference, attitude, and control). Generally, the terms were maximized roughly in unison. Each reward term had a significant increase early on in the learning process (especially in the LQR reference term), with the remainder of the learning process dedicated to making fine-tuned improvements to achieve more docks, follow the LQR reference more closely, and decrease control efforts. There is a section of learning (episodes 375,000-475,000) where the policy experiences a slight overall decline in scores, but this is corrected and improved upon by the end of the learning process. Note that the learning process is stochastic, which can produce such anomalous trends, but, in an overall sense, is successful in producing a nearly-optimized policy. 

\begin{figure}[htb!]
	\centering
	\includegraphics[width=4.25in]{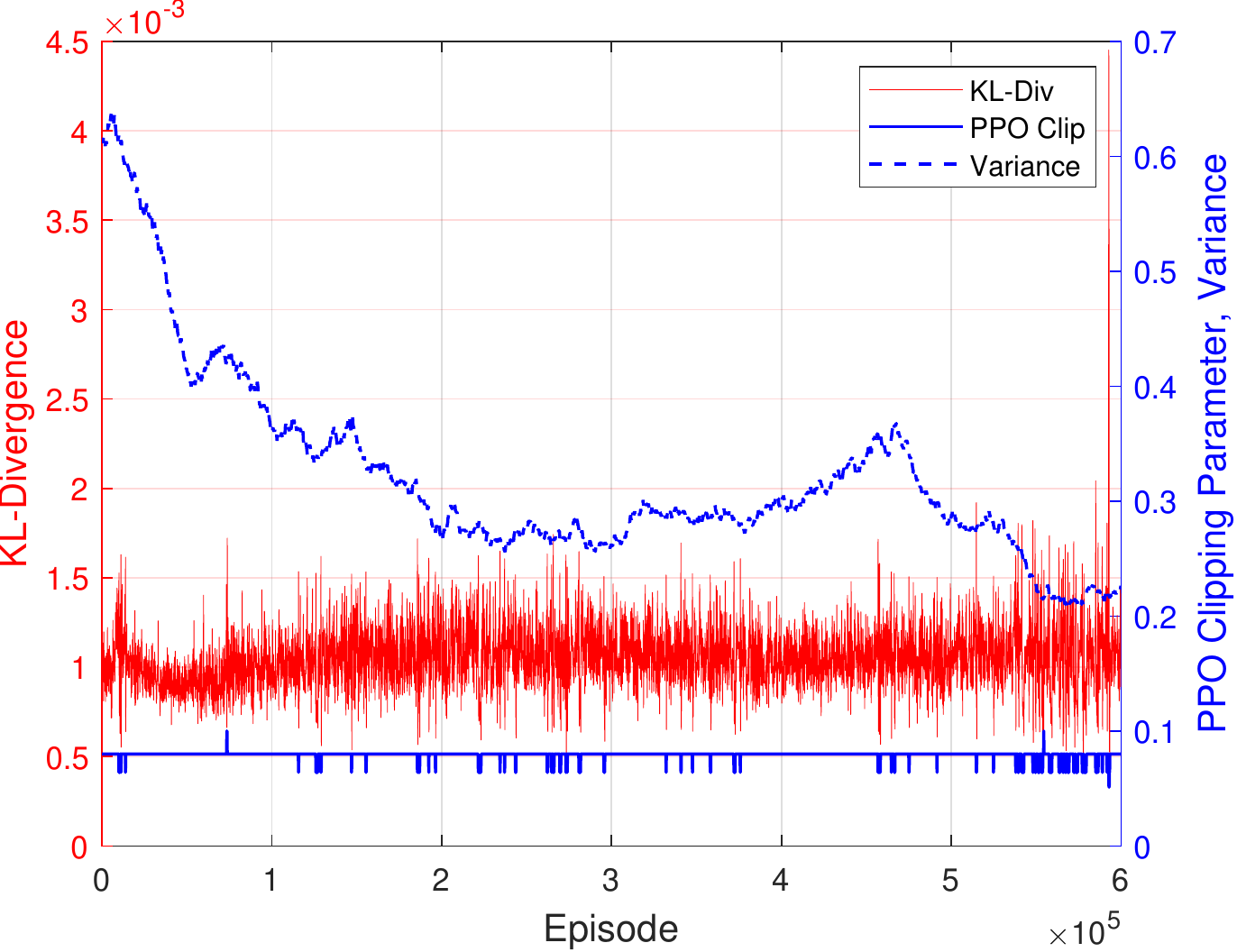}
	\caption{\textbf{KL-divergence, PPO clipping parameter $\epsilon$, and maximum action variance over the course of learning}.}
	\label{fig:kl_var}
\end{figure}

Figure \ref{fig:kl_var} depicts the KL-divergence between policy updates, the PPO clipping parameter ($\epsilon$), and the maximum action variance over the course of learning. The variance shows the desired trend: initially large to permit adequate exploration of the action space and then tapering off as the policy converges. The brief rise in variance (episodes 375,000-475,000) was a response to the slight decline in scores from Figure \ref{fig:rewards}: this permitted the policy to explore more of the action space and eventually return to an increasing score trend. The KL-divergence is largely controlled (via the adjustable PPO clipping parameter $\epsilon$ and policy learning rate $\beta_{\boldsymbol{\theta}}$) around the desired value of 0.001 (chosen to match previous RL research works) \cite{gaudet2019mars, gaudet2019asteroid}.

\subsection{Monte Carlo Test Results}
\begin{figure}[htb!]
    \centering
    \hspace*{-0.5in}
    \includegraphics[width=7in]{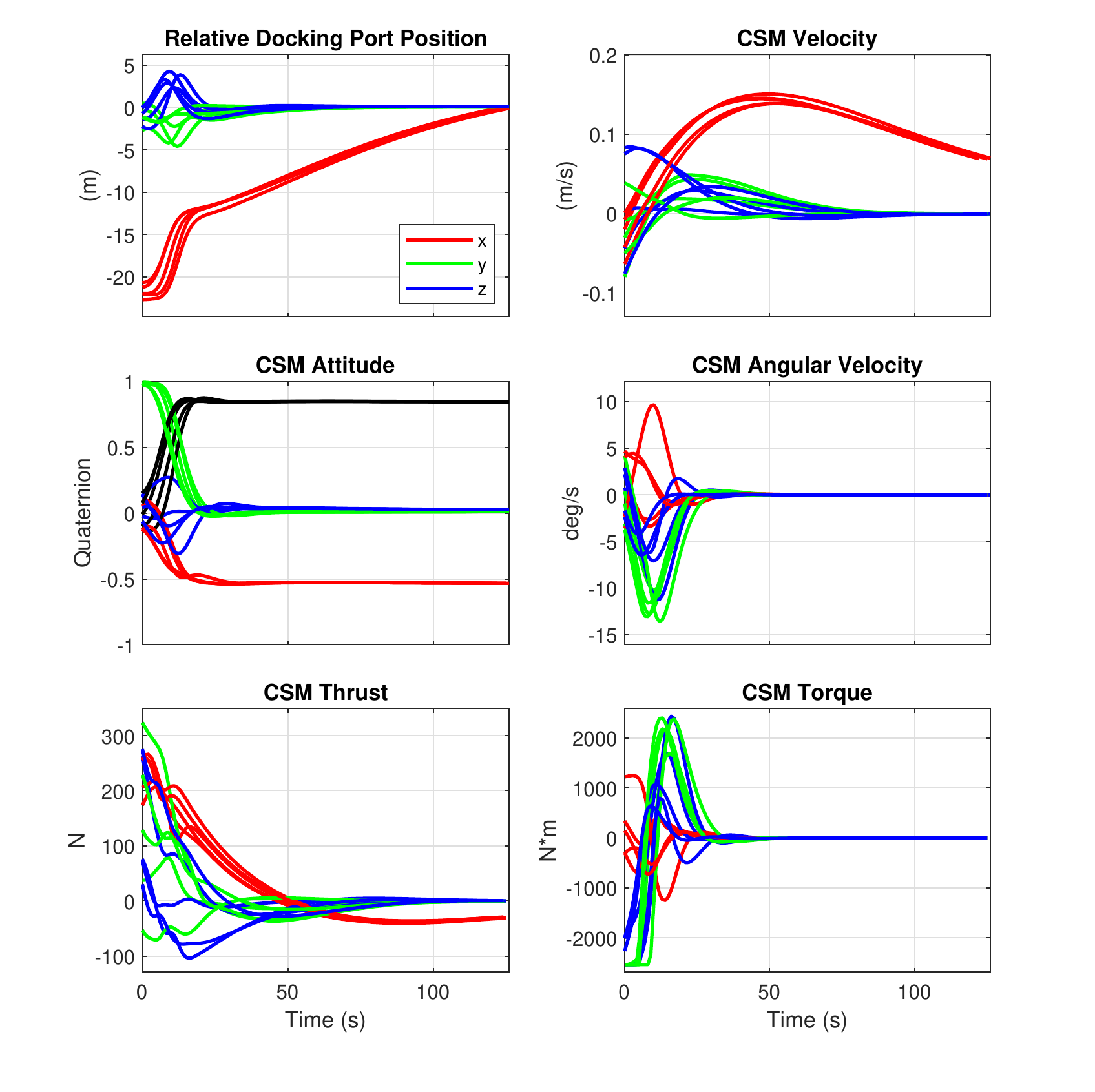}
    \vspace*{-0.35in}
    \caption{\textbf{A selection of five trajectories from the Monte Carlo test. Except for the relative docking port position, all variables are in the $\mathcal{I}$ frame.}}
    \label{fig:mctest}
    \vspace*{-0.2in}
\end{figure}

After termination of the learning process, the learned policy was tested in a series of 1000 Monte Carlo trials across randomly sampled initial conditions from the test range in Table \ref{ictable}. All variance was removed from the policy, resulting in a deterministic state-feedback control law, based on the mean policy action as a function of the state. The policy produced successful docks over all 1000 test trajectories. It took on average about 1 millisecond (on a medium performance desktop PC) for the policy neural network to compute a control input, exhibiting the fast implementation speed of an RL controller. Figure \ref{fig:mctest} shows five superimposed trajectories from the Monte Carlo trials. Table \ref{mc_data} shares key statistics on the Monte Carlo test. The total thrust and torque expenditures are calculated by integrating the thrust/torque commands over the trajectory.

Figure \ref{fig:mctest} shows that the policy exhibits varied behavior over the different trials. This points to the motivation of developing a docking policy: the resulting feedback control law produces trajectories in a fast and robust manner across a range of scenarios. Several patterns can be discerned from the Monte Carlo test. The policy produces a large negative torque in the $y$-axis even if the initial angular velocity is positive. Also, the translational position and velocity in the $y$ and $z$ axes are quickly corrected using a similar magnitude of thrust input as in the $x$-axis (attributable to the design of the LQR reference). Finally, the trajectories tend to have a slight final offset in the pitch attitude. However, note that the successful docking conditions permit small final state errors.
\begin{table}[htb!]
	\fontsize{10}{10}\selectfont
    \caption{Monte Carlo test statistics.}
   \label{mc_data}
        \centering 
      \begin{tabular}{l|l|l|c}
      \hline
      & Mean & Max & Units \\
      \hline
      Trajectory time length & $116.44$ & $135$ & s \\
      Cross-track position error [$y$, $z$] & $[0.068, 0.095]$ & $[0.085, 0.129]$ & m \\
      Cross-track velocity error [$y$, $z$] & $[7.97, 7.69]\times10^{-4}$ & $[0.0018, 0.0021]$ & m/s \\
      Final $v_x$ & $0.069$ & $0.072$ & m/s \\
      Final attitude error (axis-angle) & $5.51$ & $5.57$ & deg \\
      Final angular velocity error & $[0.013, 0.0063, 0.0041]$ & $[0.016, 0.0096, 0.012]$ & deg/s \\
      Total thrust expenditure & $8,289$ & $12,756$ & N \\
      Total torque expenditure & $54,685$ & $73,308$ & N-m \\
      \hline
   \end{tabular}
\end{table}

\subsection{Comparisons with GPOPS-II Solutions}
The General Purpose Optimal Control Software (GPOPS-II\cite{patterson2014gpops}) was utilize to provide comparisons between the converged policy and fully optimized solutions (as the former is an approximation to maximize the reward signal). The first application of GPOPS-II was to calculate the optimal trajectory using the LQR reference, attitude, and control penalties from Equation \eqref{eq:tworewards} as the objective function, thus providing a truly optimal solution to the RL reward function. The collision penalty and docking bonus terms are neglected, and instead final state constraints are included to enforce the successful docking conditions from Table \ref{tabledockconditions}. The second application of GPOPS-II was to calculate the optimal trajectory that simply minimizes overall control effort, which provides a comparison between the converged policy and the best possible solution for the problem as a whole.

\begin{figure}[htb!]
    \centering
    \hspace*{-0.5in}
    \includegraphics[width=7in]{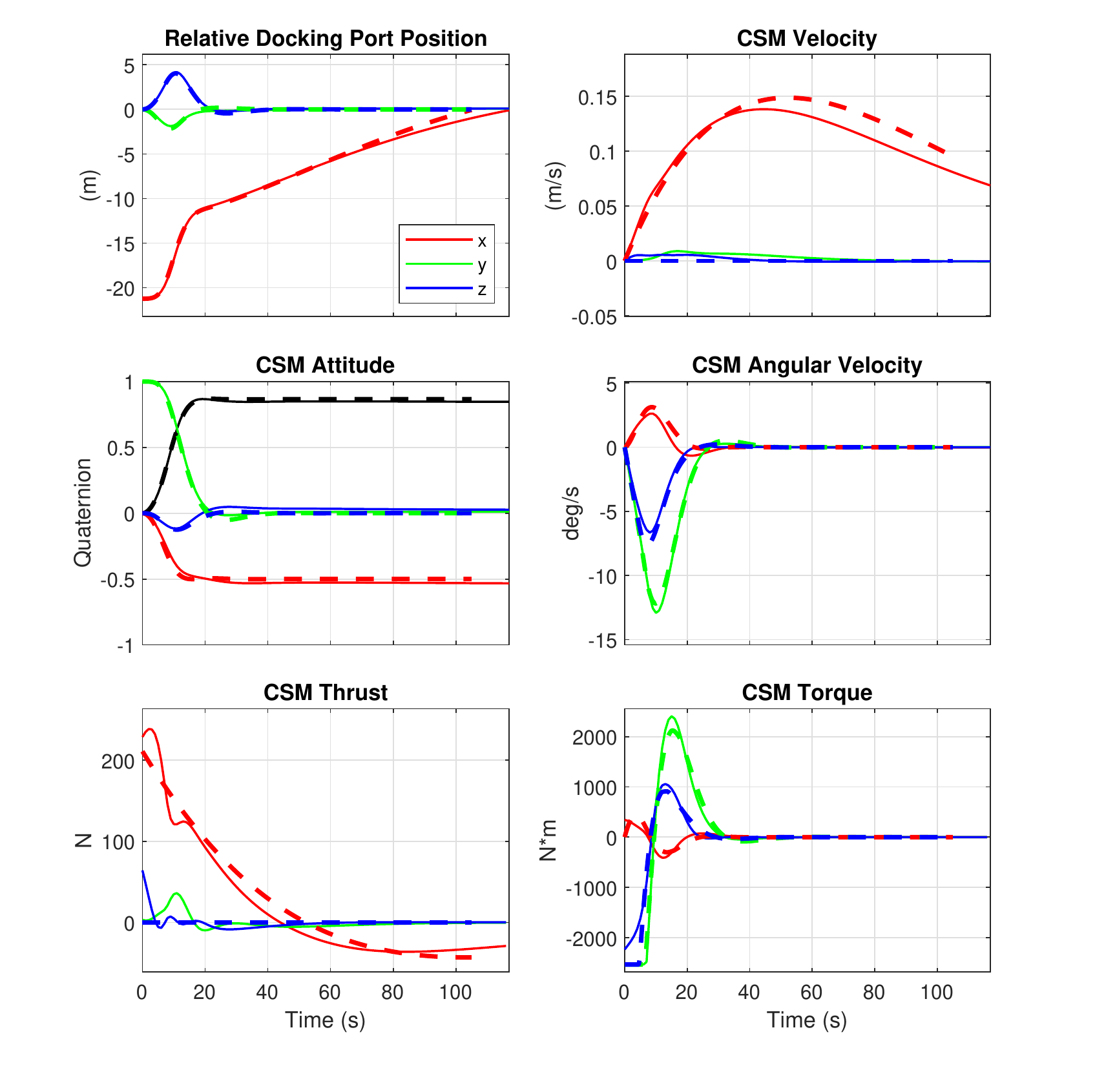}
    \vspace*{-0.35in}
    \caption{\textbf{Comparison of policy and optimal reward function trajectory (GPOPS-II) on the nominal case (dashed lines represent the GPOPS-II solution). Except the relative docking port position, all variables are in the $\mathcal{I}$ frame.}}
    \label{fig:gpopscomp}
    \vspace*{-0.2in}
\end{figure}

\begin{figure}[htb!]
    \centering
    \hspace*{-0.5in}
    \includegraphics[width=7in]{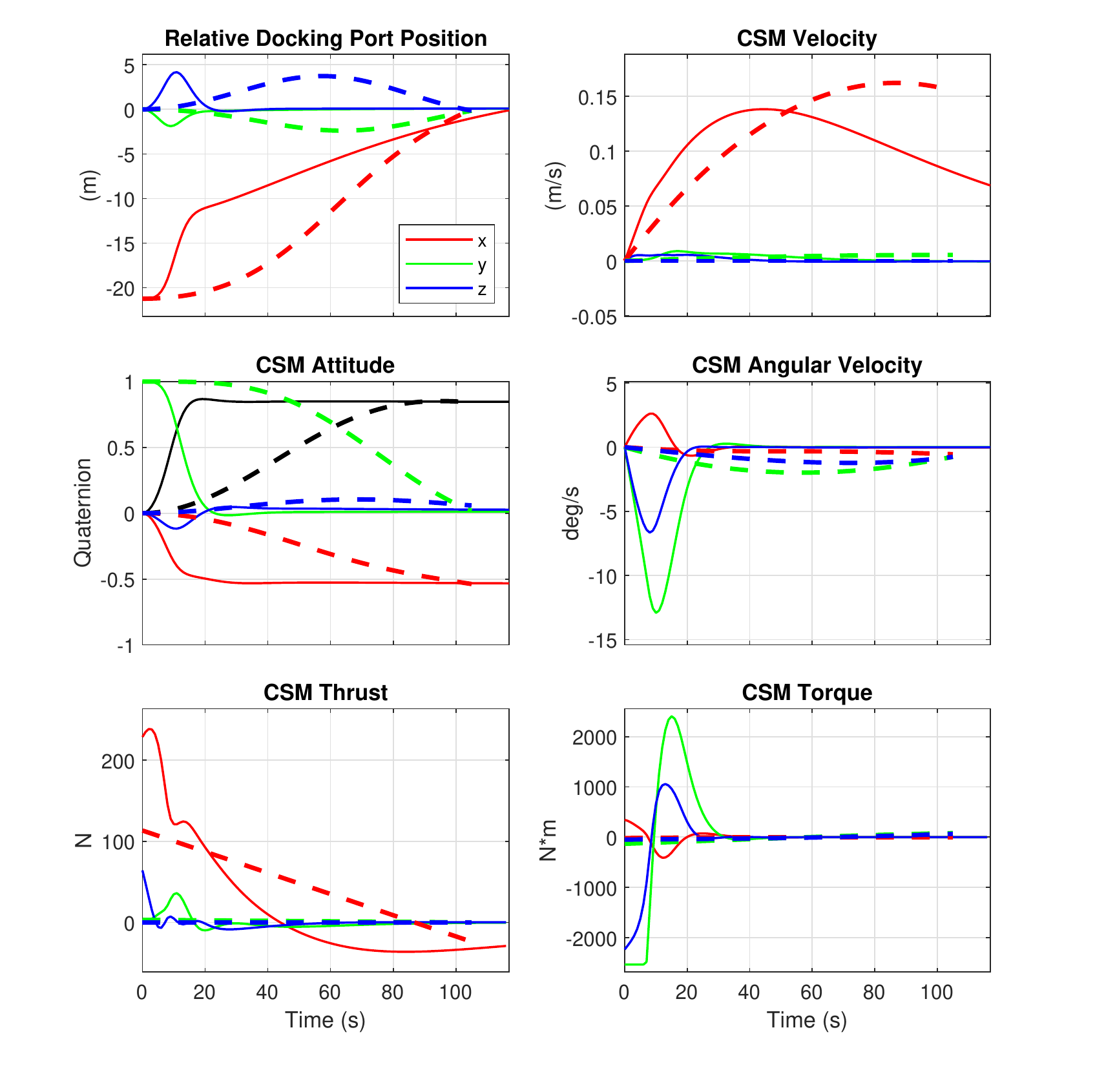}
    \vspace*{-0.35in}
    \caption{\textbf{Comparison of policy and minimum control effort trajectory (GPOPS-II) on the nominal case (dashed lines represent the GPOPS-II solution). Except the relative docking port position, all variables are in the $\mathcal{I}$ frame.}}
    \label{fig:gpopscomp_minfuel}
    \vspace*{-0.2in}
\end{figure}

The comparisons were made on the nominal docking trajectory ($\mathbf{r}_0 = [-20,~0,~0]$ m, $\mathbf{v}_0 = [0,~0,~0]$ m/s, $\mathbf{q}_0 = [0,0,1,0]$, $\boldsymbol{\omega}_0 = [0,~0,~0]$ deg/s). The GPOPS-II solutions were solved using a fixed final time of 105 seconds to match the designed LQR reference trajectory's length. The trajectory resulting from employing the policy as a closed-loop control law $\mathbf{u}_t = \pi_{\boldsymbol{\theta}}(\mathbf{x}_t)$ and the optimal trajectories produced by GPOPS-II are shown shown in Figures \ref{fig:gpopscomp}-\ref{fig:gpopscomp_minfuel}. Quantitative statistics on the comparison of the two solutions are shown in Tables \ref{tablegpopscomp}-\ref{tablegpopscomp_minfuel}.

From Figure \ref{fig:gpopscomp}, it is clear that the policy's trajectory is a (sub-optimal) approximation of the optimal trajectory for the reward function. The benefit of this approximation is that the control law $\mathbf{u}_t = \pi_{\boldsymbol{\theta}}(\mathbf{x}_t)$ resulting from the policy is quickly implementable in a closed-loop fashion (albeit within a restricted subset of the state-space). One notable difference between the two solutions is that the policy lags behind the optimal solution and produces a longer trajectory. This is likely an indicator that the policy was not fully optimized with regards to the LQR reference error penalty. Tighter adherence to the LQR reference would result in a shorter trajectory (closer to the designed 105 seconds). Also present in the policy's solution are slight, unnecessary thrust inputs in the $y$ and $z$ axes. Interestingly, the policy's solution uses slightly less overall torque effort in correcting the attitude than the optimal trajectory. This likely means that the agent more easily maximized the control reward term rather than the attitude reward term.

From Figure \ref{fig:gpopscomp_minfuel}, the policy's trajectory is noticeably different from the minimum control effort trajectory. This is largely due to the direct influence of the designed reward function on the policy's behavior. The optimal trajectory produces far less acceleration (in both translational and rotational motion) than the converged policy. Notably, to minimize control effort, the optimal trajectory results in a final state that is at the upper limit of the successful docking limits. Based on this comparison, there needs to be clear improvements to the learning process to produce a policy for minimum-fuel docking trajectories. However, simply removing the other shaping reward terms and only penalizing control effort would not be sufficient for a successful learning process. The docking problem is far too sparse to solely rely on the discrete docking bonus term for learning. As such, the other shaping penalties (LQR reference and attitude error) are needed to provide rich reward signals and improve agent performance. This comes with the disadvantage of losing control effort optimality. 

\begin{table}[htb!]
	\fontsize{10}{10}\selectfont
    \caption{Policy vs. GPOPS-II Optimal Reward Function Trajectory,  Comparison Statistics}
   \label{tablegpopscomp}
        \centering 
      \begin{tabular}{l|l|l|c}
      \hline
      & Policy & Optimal Reward Function & Units \\
      \hline
      Time Length & $117$ & $105$ & s\\
      Cross-track position error [$y$, $z$] & $[0.069, 0.095]$ & $[6.20, 5.17]\times10^{-6}$ & m \\
      Cross-track velocity error [$y$, $z$] & $[-8.04, -7.54]\times10^{-4}$ & $[-1.58,~3.06]\times10^{-6}$ & m/s \\
      Final $v_x$ & 0.083 & 0.0954 & m/s \\
      Final attitude error & $5.51$ & $1.03\times10^{-4}$ & deg \\
      Final angular velocity error & $[-0.013, -0.0063, -0.0041]$ & $[-1.81, -3.91, -2.02]\times10^{-6}$ & deg/s \\
      Thrust expenditure & $6,269$ & $6,126$ & N \\
      Torque expenditure & $53,428$ & $54,740$ & N-m \\ 
 \hline
   \end{tabular}
\end{table}
\begin{table}[htb!]
	\fontsize{10}{10}\selectfont
    \caption{Policy vs. GPOPS-II Optimal Control Effort Trajectory, Comparison Statistics}
   \label{tablegpopscomp_minfuel}
        \centering 
      \begin{tabular}{l|l|l|c}
      \hline
      & Policy & Optimal Control Effort & Units \\
      \hline
      Time Length & $117$ & $105$ & s\\
      Cross-track position error [$y$, $z$] & $[0.069, 0.095]$ & $[-0.15, -0.1436]$ & m \\
      Cross-track velocity error [$y$, $z$] & $[-8.04, -7.54]\times10^{-4}$ & $[-0.054, -0.063]$ & m/s \\
      Final $v_x$ & 0.083 & 0.15 & m/s \\
      Final attitude error & $5.51$ & $8.78$ & deg \\
      Final angular velocity error & $[-0.013, -0.0063, -0.0041]$ & $[-0.54, -0.75, -0.75]$ & deg/s \\
      Thrust expenditure & $6,269$ & $5,138$ & N \\
      Torque expenditure & $53,428$ & $6,681$ & N-m \\ 
 \hline
   \end{tabular}
\end{table}

\section{Discussion}
The motivation of using RL for 6-DOF docking is to generate a policy that is implementable as a feedback control law. This control law should be capable of producing successful docking maneuvers and be robust to initial conditions within a subset of the state-space. This contrasts with the more common, standard trajectory generation techniques (such as GPOPS-II) where the trajectory is optimized for a single scenario and must be re-calculated in the case of deviations from this particular scenario. An RL-based control law is also potentially robust to disturbances, noise, uncertain dynamics, and faults if the learning process is implemented appropriately, as exemplified in Gaudet et al. \cite{gaudet2019asteroid} 

However, there are several shortcomings with the current RL methodology, particularly with respect to docking maneuvers. First, it is difficult to target a desired time length for the docking trajectory, which may be a critical parameter in spacecraft docking operations. Secondly, it is difficult to strategically enforce constraint satisfaction. Docking scenarios often include complex path constraints such as preventing collisions, maintaining the target docking port within the sensor field-of-view, and avoiding plume impingement upon the target \cite{malyuta2019apollo}. The policy generated in this work is able to yield trajectories that suffer no collisions, but only a simple model is used and there is no \textit{explicit} guarantee that collisions are prevented. Extending the current methodology to include more complex constraints (such as plume impingement) would be challenging. Finally, the learning process as a whole is currently difficult to interpret from a design perspective, as it is sensitive to the tuning of learning and reward parameters. Thus, we expect future research efforts in the use of RL for autonomous spacecraft maneuver to address these concerns.

To facilitate future research in this area, several specific challenges experienced in this work, and their work-arounds, are discussed below.
\begin{enumerate}
    \item The dilemma of exploration vs. exploitation is a challenge in most applications of RL. In this work, the variance enables the agent to explore different control actions and better improve its maximization of rewards. However, the stochastic nature of control inputs during training due to this variance directly affects the received control effort reward signal: more variance leads to more control ``chatter'', and the agent inevitably accrues a higher overall control cost. Therefore, high control cost coefficients can possibly lead to a sharp, premature decrease in the policy variance that prohibits successful learning convergence. By carefully tuning the control cost coefficient, as well as the degree to which policy variance adjustments are made, this problem can be mitigated.
    \item The authors initially experimented with a simple state error term in the reward function to encourage the agent to achieve the desired position and velocity. However, issues were experienced with the scaling of the respective state variables and there was also no control over the trajectory time length. Thus, it is advocated to use a rich reward term \cite{gaudet2019mars, gaudet2019asteroid} that is equally effective across the entire state-space and also enables the agent to target a desired trajectory time length. In this work, the LQR reference reward term fulfills both of these objectives. However, there is certainly room for improvement as the time target was not precisely met, and the LQR design process is not valid for non-linear dynamics.
    \item Rare, but large, spikes in the KL-divergence between successive policy updates (due to the stochastic nature of action exploration) have the potential to derail the learning process. Thus, tight, frequent updates to the PPO clipping parameter and policy learning rate were made to ensure the KL-divergence stays close to the desired value of 0.001 throughout the entire learning process.
    \item In addressing the collision avoidance constraint, a smooth, continuous penalty was found to be most effective. Additionally, it is advantageous \textit{not} to terminate the episode upon collision. Terminating the episode may result in the agent determining that the most optimal policy is to violate the collision constraint as quickly as possible, end the episode, and thus avoid the accrual of other penalties in the reward function.
\end{enumerate}

\section{Conclusion}
\looseness-1
This work presents an RL framework that generates policies for autonomous 6-DOF docking maneuvers. The model-free nature of RL is appealing for developing policies that are suitable for more general classes of docking scenarios that could include significant levels of uncertainty. Using PPO, a docking policy is developed and implemented as a feedback control law over a suitably wide region of the state space, while also maintaining low on-board computational requirements. Experiments using the simulated Apollo transposition and docking maneuver validate the proposed framework and offer in-depth results regarding the learning process and test trajectory patterns. Additionally, a comparison is made between the developed policy and the solution produced by the GPOPS-II optimal control software. This comparison confirms that the RL process is able to approximate the maximization of the designed reward function with only a slight level of sub-optimality. Finally, a discussion on the current benefits and disadvantages of RL, along with highlighted implementation concerns, establishes target areas for future research. Building on these results, future work will address 6-DOF docking scenarios with uncertainty in the target's motion and the possibility of faults.

\section{Acknowledgements}
The authors would like to thank the Draper Fellow Program and the Draper Education Office for supporting this research. The authors would also like to thank the MIT Supercloud\footnote{https://supercloud.mit.edu} for providing computational resources.

\bibliographystyle{AAS_publication}   
\bibliography{references}   

\end{document}